\documentstyle[12pt]{article}

\begin{document}

\begin{titlepage}

\begin{flushright}
ENSLAPP 399/92\\
\end{flushright}
\par \vskip .1in \noindent

\begin{center}
{\LARGE  The Universal $R$-matrix and its associated\\ \par \vskip
.1in\noindent

 Quantum Algebra as functionals of the \\ \par \vskip .1in \noindent

 Classical $r$-matrix : The $sl_{2}$ case }\\
 \end{center}
  \par \vskip .3in \noindent

\begin{center}

      {\bf Laurent FREIDEL and Jean Michel MAILLET }
  \par \vskip .1in \noindent

{\sl  Laboratoire de Physique Th\'eorique $^{*}$\\
       ENS Lyon, 46 all\'ee d'Italie 69364 Lyon CEDEX 07 France}\\[0.8in]
\end{center}

\par \vskip .10in
\begin{center}
{\bf Abstract}\\
\end{center}

\begin{quote}
Using a geometrical approach to the quantum Yang-Baxter equation, the quantum
algebra ${\cal U}_{\hbar}(sl_{2})$ and its universal quantum $R$-matrix are
explicitely constructed as functionals of the associated classical $r$-matrix.
In this framework, the quantum algebra ${\cal U}_{\hbar}(sl_{2})$ is naturally
imbedded in the universal envelopping algebra of the $sl_{2}$ current algebra.
\end{quote}
\par \vskip 0.5in

\begin{flushleft}
\rule{5.1 in}{.007 in}\\
$^{*}${\small URA 1436 ENSLAPP du CNRS, associ\'ee \`a  l'Ecole Normale
Sup\'erieure de Lyon et au Laboratoire d'Annecy de Physique des Particules. }\\
{\small { email: maillet@enslapp.ens-lyon.fr}}\\
{\small { email: freidel@enslapp.ens-lyon.fr}}\\[0.2 in]

Ref. ENSLAPP 399/92 \\
September 1992
\end{flushleft}

\end{titlepage}

{\bf 1. Introduction}\\
\\

In their original construction, quantized universal envelopping algebras
appeared as the proper generalization of the corresponding Lie algebras in
which the solutions of the quantum Yang-Baxter equation do have universal
solutions (ie, not depending on a particular representation), at least as
formal power series in $\hbar$ (the Planck constant) \cite{Drin1}-\cite{FRT}.
In this context, the  link between the quantum and the associated classical
structures is the existence of a consistent classical limit $\hbar \rightarrow
0$.\\

Among several important unsolved problems in the theory of quantum algebras is
the question of universal quantization of Lie bi-algebras and of their
associated classical $r$-matrices \cite{Drin3}.\\

The purpose of this letter is to sketch a general method leading to the
construction of both quantum algebras and  their associated universal quantum
$R$-matrices as functionals of the corresponding Lie algebras and classical
$r$-matrices.\\

Our starting point is the geometrical approach to the quantum $R$-matrix
solving the quantum Yang-Baxter equation developped in \cite{JMM1} which leads
in \cite{JMM2} to a geometrical setting for quasi-triangular Hopf and
quasi-Hopf algebras associated to these quantum $R$-matrices.\\

The picture described in \cite{JMM1} is as follows. The quantum $R$-matrix can
be associated to a square surface (plaquette) and as such, can be interpreted
as a (constant) transport operator on a space of functionals of curves defined
on a                        $D$-dimensional lattice ($D \ge 3$). As shown in
\cite{JMM1}, the condition of parallel transport on this functional space is
just the quantum Yang-Baxter equation for $R$. So in this framework, the
quantum $R$-matrix plays the role of a transport operator analogous to a
Wilson-line in usual gauge theories.\\

It is then natural to ask whether there exists an analogue of the local
connection $A_{\mu} dx_{\mu}$ and of the ordered exponential mapping that
defines Wilson-lines, and what is the equation that this new local object
satisfies in order to ensure parallel transport.\\

It was argued in \cite{JMM1,JMM2} that the classical $r$-matrix defines the
local quantity analogous to $A_{\mu}$, the classical Yang-Baxter equation for
$r$ playing the role of the zero-curvature equation $F_{\mu \nu} = 0$, and that
the ordered products of $A_{\mu}$'s along a line are then replaced by  suitably
defined ordered products of $r$ on a surface.\\

In the present work we show how this picture can be realized explicitely in the
case where the initial Lie algebra is $sl_{2}$. The general case  with detailed
proofs and further applications will be given in a forthcoming publication.\\

This letter is organized as follows. In section 2 we define the local quantity
associated to the classical $r$-matrix and its surface ordered product leading
to the definition of the associated universal quantum $R$-matrix as a
functional of this classical $r$-matrix. In section 3 we apply these formulas
to the $sl_{2}$ case. We show that they lead to the construction of both the
quantum algebra ${\cal U}_{\hbar}(sl_{2})$ and its associated universal
$R$-matrix in terms of the classical $r$-matrix of $sl_{2}$, $r = \frac{1}{2} h
\otimes h + 2 e^{+} \otimes e^{-}$. In particular, the quantum generators
$S^{+}, S^{-}, H$  of ${\cal U}_{\hbar}(sl_{2})$, their algebra, and their
corresponding co-products are computed in terms of the classical one's. We give
our conclusions and comments in section 4. \\
\\

{\bf 2. The geometrical construction of the universal $R$-matrix}\\
\\

As emphasized in \cite{JMM1,JMM2}, the quantum $R$-matrix, and the Yang-Baxter
equation it satisfies, have a simple geometrical interpretation. To get an
insight into this picture (we refer for details to \cite{JMM1,JMM2}), we
consider a $D$-dimensional lattice ($D \ge 3$) with origin O and basis vectors
$e_{1}, \cdots , e_{D}$. To any elementary oriented link ($M, e_{i}$) of the
lattice, starting at point $M$ (with coordinates $(m_{1}, \cdots ,m_{D})$) in
direction $e_{i}$, we associate a Hopf algebra with unity ${\cal A}_{i}
(m_{i})$. These algebras are here considered as universal envelopping algebras
of Lie algebras ${\cal G}_{i}(m_{i})$ equiped with co-commutative co-product
${\Delta}_{0}$. All the ${\cal G}_{i}(m_{i})$ are isomorphic to a given Lie
algebra ${\cal G}$. To any oriented plaquette ($M, e_{j}, e_{i}, -e_{j},
-e_{i}$), we associate an $R$-matrix $R_{ij}(m_{i}, m_{j}) \in {\cal A}_{i}
(m_{i}) \otimes {\cal A}_{j} (m_{j})$. Note that ${\cal A}_{i} (m_{i})$ is
invariant under translations of th

e point $M$ along any vector $e_{j},\ j \neq i$. Similarly, the quantum
$R$-matrix $R_{ij}(m_{i}, m_{j})$ is invariant under translations of the point
$M$ along any vector $e_{k},\ k \neq i,\ k \neq j$. A particular case of this
picture is given by putting the same $R$-matrix on all the plaquettes of the
lattice. The local version of the quantum Yang-Baxter equation for these
$R$-matrices reads (see \cite{JMM1}),
\begin{equation}
R_{ij}(m, n) \ R_{ik}(m, p) \ R_{jk}(n, p)\ =\ R_{jk}(n, p) \ R_{ik}(m, p) \
R_{ij}(m, n)
\label{eq:qyb}
\end{equation}
with $i, j, k \in \{ 1, \cdots , D \}$ and $m, n, p \in {\bf Z}$. It reduces to
the usual quantum Yang-Baxter equation in the above particular case.\\

The geometrical meaning of this equation is the following.\\
Let ($M, e_{k}, e_{j}, e_{i}$) be a cube drawn on the lattice, and let us
divide its boundary into two parts having as common boundary the closed
oriented curve $C = (M, e_{k}, e_{j}, e_{i}, - e_{k}, - e_{j}, - e_{i})$. To
each of these two surfaces we associate an ordered product of three
$R$-matrices (corresponding to the three respective plaquettes they are build
of) which are precisely given by the right and left hand sides of eq.
(\ref{eq:qyb}). Hence, the quantum Yang-Baxter equation (\ref{eq:qyb}) ensures
the equality of operators associated to different surfaces having the same
oriented boundary. This picture acquires a natural interpretation as a zero
curvature equation on a space of functional of curves \cite{JMM1}.\\

We now define the notion of surface ordered products of $R$-matrices for plane
rectangular surfaces $\Gamma_{ij} (M, N)$ starting at point $O$ in the plane
$(e_{i}, e_{j})$, and bounded by the oriented curve ($O, Ne_{j}, Me_{i}, -
Ne_{j}, - Me_{i}$). We associate to such a surface the following ordered
product of the $R$-matrices,
\begin{eqnarray}
R_{{\hat i}, {\hat j}}^{(M, N)}\ &=&\  R_{{\hat i}, j}^{(M)}(N-1) \  R_{{\hat
i}, j}^{(M)}(N-2) \cdots  R_{{\hat i}, j}^{(M)}(0),\nonumber\\
&=&\ R_{i, {\hat j}}^{(N)}(0) \  R_{i, {\hat j}}^{(N)}(1)  \cdots  R_{i, {\hat
j}}^{(N)}(M-1)
\label{eq:RMN}
\end{eqnarray}
where we have defined
\begin{eqnarray}
R_{{\hat i}, j}^{(M)}(k)\ &=&\  R_{ij}(0, k) \  R_{ij}(1, k) \cdots
R_{ij}(M-1, k),\nonumber\\
R_{i, {\hat j}}^{(N)}(k) &=&  R_{ij}(k, N-1)\ R_{ij}(k, N-2) \cdots R_{ij}(k,
0)
\label{eq:RNRM}
\end{eqnarray}
The above surface ordered product of $R$-matrices implements the geometrical
gluing of the plaquettes building up $\Gamma_{ij} (M, N)$.
It is rather easy to prove, both geometrically and algebraically, that if the
quantum Yang-Baxter equations (\ref{eq:qyb}) are satisfied, then $R_{{\hat i},
{\hat j}}^{(M, N)}$ also satisfies it, namely,
\begin{equation}
R_{{\hat i}, {\hat j}}^{(M, N)}\ R_{{\hat i}, {\hat k}}^{(M, P)}\ R_{{\hat j},
{\hat k}}^{(N, P)}\ =\ R_{{\hat j}, {\hat k}}^{(N, P)}\ R_{{\hat i}, {\hat
k}}^{(M, P)}\ R_{{\hat i}, {\hat j}}^{(M, N)}
\label{eq:bqyb}
\end{equation}
Note that the order of products of $R$-matrices in eqs. (\ref{eq:RMN},
\ref{eq:RNRM}) is according to increasing $m$ indices in first space $(i)$,
while it is in opposite order in the second space $(j)$.\\

In these notations, we define a local quantum co-product law
$$
{\Delta}\ :\ {\cal A}_{i}(m) \rightarrow {\cal A}_{i}(m) \otimes {\cal
A}_{i}(m+1)
$$
so that its action on the above $R$-matrices is given as the operator
associated to rectangles obtained by making the geometrical gluing of two
adjacent squares, \begin{eqnarray}
(\Delta \otimes {\bf 1}) R_{ij}(m, n)\ &=&\ R_{ij}(m, n)\ R_{ij}(m+1,
n)\nonumber\\
({\bf 1} \otimes \Delta) R_{ij}(m, n)\ &=&\ R_{ij}(m, n+1)\ R_{ij}(m, n)
\label{eq:deltar}
\end{eqnarray}
Note that such $R$-matrices are quasi-triangular. These co-product actions are
compatible with the above surface ordering, namely we have,
$$
R_{{\hat i}, {\hat j}}^{(M, N)}\ =\ ({\Delta}^{(M)} \otimes  {\Delta}^{(N)})
R_{ij}(0, 0)
$$

We now describe the way to compute the universal $R$-matrix as a functional of
its classical counterpart.\\

We first take the lattice spacing $(a)$ as a small parameter so that in the
limit $a \rightarrow 0$ the quantum $R$-matrix $R_{ij}(m, n)$ admits the
expansion,
\begin{equation}
R_{ij}(m, n)\ =\ {\bf 1}\ +\ \hbar a^2 \ r_{ij}(m, n)\ +\ O(\hbar^{2} a^{4})
\label{eq:Rr}
\end{equation}
where the classical $r$-matrix $r_{ij}(m, n)$ takes values in ${\cal G}_{i}(m)
\otimes {\cal G}_{j}(n)$, $i, j \in \{1, \cdots, D\}$, $m, n \in {\bf Z}$, and
${\cal G}_{i}(m)$ are Lie algebras isomorphic to a given Lie algebra $\cal G$.
More precisely, let the classical $r$-matrix be an element of $\cal G \otimes
\cal G$ defined by (unless necessary, and for the sake of simplicity we
suppress in the following the indices $(ij)$ on the $r$-matrix),
$$
r\ =\ r^{\alpha \beta}\ E_{\alpha}\ \otimes\ E_{\beta}
$$
with the defining Lie algebra relations,
$$
[\ E_{\alpha},\ E_{\beta}\ ]\ =\ C_{\alpha \beta}^{\gamma}\ E_{\gamma}
$$
Then the local $r$-matrix $r(m, n)$ is defined by,
$$
r(m, n)\ =\ r^{\alpha \beta}\ E_{\alpha}(m)\ \otimes\ E_{\beta}(n)
$$
with the commutation relations (for the local Lie algebras ${\cal G}(m)$),
$$
[\ E_{\alpha}(m),\ E_{\beta}(n)\ ]\ =\ a^{-1} \delta_{mn} C_{\alpha
\beta}^{\gamma}\ E_{\gamma}(m)
$$

The  universal quantum $R$-matrix $\cal R$ associated to a given classical
$r$-matrix is obtained as the limit $N \rightarrow \infty$ of the square
$R$-matrix $R_{{\hat i}, {\hat j}}^{(N, N)}$ while keeping the product $a \cdot
N$ fixed, say $a \cdot N\ =\ 1$ \footnote{\em This procedure is analogous to
the construction of the ordered exponential integral of a connection $A_{\mu}
\in {\cal G}$ along a line as the limit $N \rightarrow \infty$ of the line
ordered product of $N$ objects. Here we are dealing with the analogous
construction but for $r$-type objects, $r \in {\cal G} \otimes {\cal G}$ and
for surfaces.}. In this limit $R_{{\hat i}, {\hat j}}^{(N, N)}$ will take
values in the tensor product $\otimes_{m=0}^{N-1} {\cal A}_{i} (m)
\otimes_{n=0}^{N-1}{\cal A}_{j} (n)$ with $N \rightarrow \infty$, and it is
convenient to relabel the integer indices $m, n$ by $u\ =\ m a$ and $v\ =\ n
a$, $u, v$ becoming real numbers in $[0, 1]$. Using these notations, $r(m, n)$
is to be replaced by $r(u,~v)$, and $E_{\al

pha}(m)$ by $E_{\alpha}(u)$.\\

Thus, in the limit $a \rightarrow 0$, $r(u, v)$ is taking values in ${\hat
{\cal G}} \otimes {\hat {\cal G}}$ where ${\hat {\cal G}}$ is the current
algebra over $\cal G$ defined by the commutation relations,
\begin{equation}
[\ E_{\alpha}(u),\ E_{\beta}(v)\ ]\ =\ \delta(u-v)\ C_{\alpha \beta}^{\gamma}\
E_{\gamma}(u)
\label{eq:ca}
\end{equation}
with $u, v \in {\bf R}$, and
\begin{equation}
r(u, v)\ =\ r^{\alpha \beta}\ E_{\alpha}(u)\ \otimes\ E_{\beta}(v)
\label{eq:ruv}
\end{equation}

The next step is to extract from the above picture a compact formula for the
universal quantum $R$-matrix $\cal R$ as a functional of the classical
$r$-matrix. To achieve this we first notice that the ordered product
(\ref{eq:RMN}) defining  $R^{(N)}$ can be recasted in the following way,
\begin{equation}
R^{(N)}\ =\ R^{(N)} [-N+1]\ R^{(N)} [-N+2]\ \cdots\ R^{(N)} [N-1]
\label{eq:RN}
\end{equation}
where,
\begin{equation}
R^{(N)} [p]\ =\ \prod_{p+n-m = 0}\ R(m, n)
\label{eq:Rp}
\end{equation}
$n,\ m\ \in\  \{1, \cdots N\}$ and $p\ \in\ \{-N+1,\ \cdots\ N-1 \}$. Note that
in $R^{(N)} [p]$ all $R(m, n)$ are commuting among themselves. In the limit $a
\rightarrow 0$  we obtain,
\begin{equation}
R^{(N)} [p]\ =\ {\bf 1}\ +\ \hbar a\ r^{(N)} [p] \ +\ O(\hbar^{2} a^{2})
\label{eq:rp}
\end{equation}
with
$$
r^{(N)} [p]\ =\ a \sum_{p+n-m = 0}\ r(m, n)
$$
\\
Hence in the limit $N \rightarrow \infty$, the above ordered product
(\ref{eq:RN}) becomes an ordered exponential integral of the potential
$r[\omega]\ =\ \lim_{N \rightarrow \infty} r^{(N)} [p]$, ($\omega\ =\ p \cdot
a$), defined as,
\begin{equation}
r [\omega]\ =\ \int^{1}_{0} \int^{1}_{0}\ du dv\ r(u, v)\ \delta( \omega + v -
u)
\label{eq:rw}
\end{equation}
Note that this operator $r [\omega]$ as compact support, namely it is zero if
$| \omega | \ge 1$. Moreover all elements $r(u,v)$ in eq. (\ref{eq:rw}) are
comuting among themselves for a given $\omega$. We finally get for the
universal quantum $R$-matrix ${\cal R}\ =\ \lim_{N \rightarrow \infty}\
R^{(N)}$,
\begin{equation}
{\cal R}_{12}\ =\ \overrightarrow{\exp} \ ( \hbar \int_{-1}^{+1} d{\omega}\
r_{12}[\omega] )
\label{eq:RU}
\end{equation}

This formula defines the surface ordered product of the classical $r$-matrix on
a square of size $1$. Note that the classical $r$-matrix entering this formula
doesn't need to be skew-symmetric. From the above definition, $\cal R$ is an
element of ${\cal U}(\hat {\cal G}) \otimes {\cal U}(\hat {\cal G})$, ${\cal
U}(\hat {\cal G})$ being the universal envelopping algebra of the current
algebra ${\hat {\cal G}}$ over the Lie algebra $\cal G$
\footnote{\em  In fact we consider the completion of ${\cal U}(\hat {\cal G})$
containing formal power series in $\hbar$ with coefficients of the type $\int
du_{1} \cdots du_{n}\ e_{{\alpha}_{1}}(u_{1})\cdots e_{{\alpha}_{n}}(u_{n})\
{\varphi}(u_{1},\cdots,u_{n})$, where the $\varphi$'s have compact support.}.
However, note that in general $\cal R$ is not an element of  ${\cal U}({\cal
G}) \otimes {\cal U}({\cal G})$.\\
The following fundamental property is verified,\\

{\em Let the classical $r$-matrix satisfy the classical Yang-Baxter equation,}
$$
\left[\ r_{12}\ ,\ r_{13}\ \right]\ +\ \left[\ r_{12}\ ,\ r_{23}\ \right]\ +\
\left[\ r_{13}\ ,\ r_{23}\ \right]\ =\ 0
$$
{\em then, the operator $r [\omega]$ satisfies the classical Yang-Baxter
equation with additive spectral parameter $\omega$, and the operator $\cal R$
defined in (\ref{eq:RU}) is invertible and satisfies the quantum Yang-Baxter
equation}
$$
{\cal R}_{12}\ {\cal R}_{13}\ {\cal R}_{23}\ =\ {\cal R}_{23}\ {\cal R}_{13}\
{\cal R}_{12}
$$
\\

It is interesting to note that analogous (but not identical) formulas were
obtained for the quantum $R$-matrix in particular matrix representations of
${\cal G}$ in \cite{BAZ1} and in \cite{TAK1} from a completely different point
of view.\\

We also define the surface ordered product of $r(u, v)$ on the triangles
defined respectively by $\omega \in [0, 1]$ and $\omega \in [-1, 0]$ as,
\begin{equation}
{\cal F}_{12}\ =\ \overrightarrow{\exp} \ ( \hbar \int_{0}^{1} d{\omega}\
r_{12} [\omega] )
\label{eq:f}
\end{equation}
and similarly
\begin{equation}
{\tilde {\cal F}}_{12}\ =\ \overrightarrow{\exp} \ ( \hbar \int_{-1}^{0}
d{\omega}\ r_{12} [\omega] )
\label{eq:ft}
\end{equation}
We denote ${\cal F}_{12}$ by ${\cal F}^{(+)}_{12}$ and by ${\cal F}^{(-)}_{12}$
the object $\cal F$ obtained from eq. (\ref{eq:f}) with $r_{12}\ \equiv\
r^{(+)}_{12}$ replaced by  $r^{(-)}_{12}\ \equiv\  - {\sigma}( r_{12})$,
${\sigma}$ being the usual transposition operator in ${\cal G} \otimes {\cal
G}$, ${\sigma}(x \otimes y)\ =\ y \otimes x$.\\
In a similar way we define ${\cal R}^{(\pm)}_{12}$ from eq. (\ref{eq:RU}). We
get,
${\tilde {\cal F}}_{12}\ =\ {\cal F}^{(-) -1}_{21}$, hence we have the
property,\\

{\em The universal quantum $R$-matrix factorizes as,}
$$
{\cal R}^{(+)}_{12}\ =\ {\cal F}^{(-) -1}_{21}\ {\cal F}^{(+)}_{12}
$$
{\em and}
$$
{\cal R}^{(-)}_{12}\ =\ {\cal R}^{(+)-1}_{21}
$$
{\em satisfies the quantum Yang-Baxter equation if ${\cal R}^{(+)}$ does. If
the classical $r$-matrix is skew-symmetric, namely $r^{(+)}\ =\ r^{(-)}$, then
${\cal F}^{(+)}_{12}\ =\ {\cal F}^{(-)}_{12}\ =\ {\cal F}_{12}$ and the quantum
$R$-matrix $\cal R$ becomes of triangular type,}
$$
{\cal R}_{12}\ =\ {\cal F}^{-1}_{21}\ {\cal F}_{12}
$$
\\
This decomposition of ${\cal R}$ implements the geometrical decomposition of a
square into two triangles.\\

These $\cal F$ operators are in a sense more fundamental than the universal
quantum $R$-matrix $\cal R$. Indeed, not only $\cal R$ can be obtained from
them, but they also give the deformation sending the co-commutative co-product
$\Delta_{0}$ in  ${\cal U}(\hat {\cal G})$, (which is the one coming from the
original Lie algebra $\cal G$), to the quantum co-product $\Delta$ associated
to $\cal R$. The following property is satisfied,\\

{\em Let the classical $r$-matrix satisfy the classical Yang-Baxter equation,
then, the co-products defined for any element  $a \in {\cal U}(\hat {\cal G})$
by,}
$$
{\Delta}^{(\pm)} (a)\ =\ {\cal F}^{(\pm)-1}_{12}\ {\Delta}_{0}(a)\ {\cal
F}^{(\pm)}_{12}
$$
{\em are co-associatives. They give ${\cal U}(\hat {\cal G})$ the structure of
a Hopf algebra with quasi-triangular universal quantum $R$-matrix $\cal R$
satisfying the properties,}
$$
{\cal R}\ {\Delta}^{(+)} (a)\ =\ {\sigma} \circ {\Delta}^{(-)} (a) \ {\cal R}
$$
$$
({\Delta}^{(\pm)} \otimes {\bf 1})\ {\cal R}\ =\ {\cal R}_{13}\ {\cal R}_{23}
$$
$$
({\bf 1} \otimes {\Delta}^{(\pm)})\ {\cal R}\ =\ {\cal R}_{13}\ {\cal R}_{12}
$$
{\em $\sigma$ being the usual transposition operator in ${\cal U}(\hat {\cal
G}) \otimes {\cal U}(\hat {\cal G})$.}\\

Note that the co-products ${\Delta}^{(+)}$ and ${\Delta}^{(-)}$ are equal when
they are acting on $\cal R$. Hence the quantum Yang-Baxter equation for $\cal
R$ follows from the above property. It also gives the possibility of defining a
quasi-triangular Hopf sub-algebra of ${\cal U}(\hat {\cal G})$ with induced
co-product ${\Delta}\ =\ {\Delta}^{(+)}\ =\ {\Delta}^{(-)}$.\\

The co-associativity of ${\Delta}^{(\pm)}$ follows from the co-associativity of
${\Delta}_{0}$ and  the formulas
\begin{equation}
({\Delta}_{0} \otimes {\bf 1}) {\cal F}^{(\pm)}\ {\cal F}^{(\pm)}_{12}\ =\
({\bf 1} \otimes {\Delta}_{0}) {\cal F}^{(\pm)}\ {\cal F}^{(\pm)}_{23}
\end{equation}
which are direct consequences of the classical Yang-Baxter equation for the
classical $r$-matrix and the obvious ${\Delta}_{0}$ actions,
\begin{eqnarray}
({\Delta}_{0} \otimes {\bf 1}) {\cal F}\ &=&\ \overrightarrow{\exp} \ ( \hbar
\int_{0}^{+1} d{\omega}\ (r_{13}[\omega]\ +\ r_{23}[\omega]))\nonumber\\
({\bf 1} \otimes {\Delta}_{0}) {\cal F}\ &=&\  \overrightarrow{\exp} \ ( \hbar
\int_{0}^{+1} d{\omega}\ (r_{12}[\omega]\ +\ r_{13}[\omega]))
\end{eqnarray}\\
\\

{\bf 3. The Geometrical construction of ${\cal U}_{\hbar}(sl_{2})$}\\
\\

The Lie algebra $sl_{2}$ is generated by the elements $h,\ e^{+},\ e^{-}$
satisfying the relations,
\begin{eqnarray*}
\left[h, e^{\pm}\right] &=& \pm 2 e^{\pm} \nonumber\\
\left[e^{+}, e^{-}\right] &=& h
\end{eqnarray*}
The classical $r$-matrix associated to $sl_{2}$ is given by,
$$
r = \frac{1}{2} h \otimes h + 2 e^{+} \otimes e^{-}
$$
It satisfies the classical Yang-Baxter equation. Note it is not skew-symmetric.
The $sl_{2}$  current algebra is defined as in eq.(\ref{eq:ca}).\\

We now apply the general procedure of the preceding section to this simple
case. We will use the following notations. To any element $a(u,v) \in
\widehat{sl_{2}} \otimes \widehat{sl_{2}}$
we associate $a\left[\omega\right]$ defined by,
$$
a\left[\omega\right]=\int_{0}^{1}\ du \int_{0}^{1}\ dv\ a(u,v)\
\delta(u-v-\omega)
$$
$a\left[\omega\right]$ has a compact support, namely, $a\left[\omega\right]
=0\: if\: |\omega|> 1$.
We now compute the quantum universal $R$-matrix according to eq. (\ref{eq:RU}).
It can be recasted in the following form,
\begin{eqnarray}
{\cal R} & = & \overrightarrow{\exp}\ (\hbar \int_{-1}^{+1}d\omega\
r\left[\omega\right])\nonumber\\
& = & \overrightarrow{\exp}\ ( \hbar \int_{-1}^{+1} d\omega\ \left[ \frac{1}{2}
 (h \otimes h)\left[\omega\right] + 2 (e^{+} \otimes e^{-}) \left[\omega
\right] \right] ) \nonumber\\
& = & \overrightarrow{\exp}\ ( \frac{\hbar}{2}\int_{-1}^{+1} d\omega\  ( h
\otimes h) \left[ \omega \right])\ \overrightarrow{\exp}\ ( 2
\hbar\int_{-1}^{+1}\ d\omega\ (S^{+} \otimes S^{-}) \left[ \omega \right] )
\end{eqnarray}

\begin{eqnarray}
S^{+}(u) & = & e^{+}(u)\ q^{\int_{u}^{1} dv\ h(v) }\nonumber\\
S^{-}(u) & = & q^{- \int_{0}^{u} dv\ h(v)}\ e^{-}(u) \nonumber\\
q & = & \exp{(\hbar)}
\end{eqnarray}
To prove this, we use the $sl_{2}$ current algebra and the following identities
for ordered exponential integrals,
\[ \overrightarrow{\exp}\ ( \int_{a}^{b} d\omega\ \left[ A(\omega) + B(\omega)
\right] )\ =\ \overrightarrow{\exp}\ ( \int_{a}^{b}d\omega\ A(\omega) )\ \
\overrightarrow {\exp}\ ( \int_{a}^{b} d\omega\ B_{A}(\omega) ) \]
with \[ B_{A}(\omega)\ =\ \left[ \overrightarrow{\exp}\ ( \int_{\omega}^{b}
d\omega'\ A(\omega') )\right]^{-1}  B(\omega) \  \left[\overrightarrow{\exp}\ (
\int_{\omega}^{b} d\omega'\ A(\omega') ) \right] \]
We obtain the (deformed) current algebra,
\begin{eqnarray}
S^{+}(u)\ S^{+}(v) & = & q^{2\epsilon(v-u)}\ S^{+}(v)\ S^{+}(u) \nonumber\\
S^{-}(u)\ S^{-}(v) & = & q^{2\epsilon(u-v)}\ S^{-}(v)\ S^{-}(u) \nonumber\\
\left[S^{+}(u)\ ,\ S^{-}(u')\right] & = & {(-2 \hbar)}^{-1}\ ({\partial}_{u}
K(u))\  \delta(u-u')\nonumber\\
K(u)\ S^{(\pm)}(v) & = & q^{{\pm}2\epsilon(v-u)}\ S^{(\pm)}(v)\ K(u)\nonumber\\
K(u) & = & q^{\int_{u}^{1} dv\ h(v)\  -\ \int_{0}^{u} dv\ h(v) }
\end{eqnarray}
We now proceed to compute the above ordered exponential integrals. The one
containing $h$'s only can be symmetrized quite easily since all $h(u)$ are
commuting among themselves. For the other one we get,
\[ \overrightarrow{\exp}\ ( 2 \hbar\int_{-1}^{+1}\ d\omega\ (S^{+} \otimes
S^{-}) \left[ \omega \right] )\ =\ 1 + \sum_{n\geq 1} (2{\hbar})^{n} S_{n} \]
with \[ S_{n}= \int_{0}^{1} \prod_{i =1}^{n}du_{i}dv_{i}\left(\prod_{i
=1}^{n-1} \theta(\omega_{i+1}-\omega_{i})\right) {\cal S}(u_{1},v_{1})\dots
{\cal S} (u_{n},v_{n})  \]
\begin{eqnarray*}
& & \omega_{i} =u_{i}-v_{i} \\
& & {\cal S}(u,v) = S^{+}(u) \otimes S^{-}(v) \hspace{6cm} \\
& &\theta(x) = \left\{\begin{array}{ll}
 0 & \mbox{if $x\leq0$} \\
+1 & \mbox{if $x>0$} \end{array} \right.
\end{eqnarray*}
To get rid of the theta distributions in the integrant we first remark that \\
$\theta(\omega_{j}-\omega_{i})\ \theta(u_{i}-u_{j})\ \left[ {\cal
S}(u_{i},v_{i})\,\ {\cal S}(u_{j},v_{j}) \right] =0$. Then as the commutation
relations of $\cal S$ produce only $\theta$ distributions, we can use the
identity $1 = \sum_{\sigma \in S_{n}}\prod_{i =1}^{n-1} \theta(u_{\sigma(i)}
-u_{\sigma(i+1)} )$ in the integrant and transport the permutation operations
on the variables $\omega$. Hence we obtain,
\begin{eqnarray}
S_{n}&=&\sum_{\sigma \in S_{n}} \int_{0}^{1} \prod_{i=1}^{n}du_{i}dv_{i}
\left(\prod_{i =1}^{n-1} \theta(\omega_{\sigma_{(i+1)}}-\omega_{\sigma_{(i)}})
\right)\left(\prod_{j =1}^{n-1} \theta(u_{j} -u_{j+1} )  \right)\nonumber\\
\nonumber\\
&\cdot&{\cal S}(u_{\sigma_{(1)}},v_{\sigma_{(1)}}) \dots  {\cal
S}(u_{\sigma_{(n)}},v_{\sigma_{(n)}})
\end{eqnarray}
Suppose that $n = \sigma_{(i)}$ then for all $j<i$ there exists a distribution
$\theta(\omega_{n}-\omega_{\sigma(j)})$ in the corresponding term.\\
Moreover for all $j \neq i$  $\sigma_{(j)}<n$ so there also exists a
distribution  $\theta(u_{\sigma(j)}-u_{n})$ in the corresponding term. Thus we
can commute ${\cal S}(u_{n},v_{n}) $ to the left. Applying this argument
recursively we  finally reorder all the products ${\cal S}(u_{\sigma_{(1)}},
v_{\sigma_{(1)}}) \dots {\cal S}(u_{\sigma_{(n)}}, v_{\sigma_{(n)}}) $ into $
{\cal S}(u_{n},v_{n})  \dots  {\cal S}(u_{1},v_{1}) $ and reconstruct $1$ as
the sum of products of distributions $\theta$ to give,
\begin{eqnarray}
S_{n}&=&\int_{0}^{1} \prod_{i =1}^{n}du_{i} dv_{i}\left(\sum_{\sigma \in S_{n}}
\prod_{i =1}^{n-1} \theta(\omega_{\sigma_{(i+1)}}-\omega_{\sigma_{(i)}})
\right) \left(\prod_{j =1}^{n-1} \theta(u_{j} -u_{j+1} )  \right)\nonumber\\
\nonumber\\
&\cdot&{\cal S}(u_{n},v_{n}) \dots  {\cal S}(u_{1},v_{1})\nonumber\\
\nonumber\\
&=&\int_{0}^{1} \prod_{i =1}^{n}du_{i}\left(\prod_{j =1}^{n-1} \theta(u_{j}
-u_{j+1} )  \right) S^{+}(u_{n}) \dots S^{+}(u_{1}) \otimes \left(
\int_{0}^{1}{dv} S^{-}(v) \right)^{n}\nonumber\\
\end{eqnarray}
We also have,
\begin{equation}
\int_{0}^{1} \prod_{i =1}^{n}du_{i}\left(\prod_{j =1}^{n-1} \theta(u_{j}
-u_{j+1} )  \right) S^{+}(u_{n}) \dots S^{+}(u_{1}) =
\frac{1}{\left\{n\right\}_{q} !} \left(\int_{0}^{1}{du} S^{+}(u) \right)^{n}
\end{equation}
with
\[ \left\{n\right\}_{q} = \frac{1-q^{-2n}}{1-q^{-2}} = q^{-n+1} \left[ n
\right]_{q} \]
We define,
$$
 \exp_{q}(x) = \sum_{n=0}^{\infty} { \frac{x^{n}}{\left\{n\right\}_{q}!}  }
$$
It verifies,
$$
\exp_{q}(x+y) = \exp_{q}(x)\exp_{q}(y) \mbox{ if }  xy = q^{2}yx
$$
Finally putting all these formulas together we obtain,
\begin{eqnarray}
{\cal R} & = & \overrightarrow{\exp}\ ( \hbar \int_{-1}^{+1} d\omega\
r\left[\omega\right] )\nonumber\\
  & = & q^{\frac{1}{2}H\otimes H }\ {\exp}_q(2\hbar \ S^{+} \otimes S^{-})
\end {eqnarray}
\begin{eqnarray*}
\mbox{where}&  S^{+} = \int_{0}^{1} du\ S^{+}(u) & \\
            &  S^{-} = \int_{0}^{1} du\ S^{-}(u) & \\
            &   H   =  \int_{0}^{1} du\ h(u) &
\end{eqnarray*}
satisfy the following ${\cal U}_{\hbar}(sl_{2})$-commutation relations
\footnote{\em In the semi-classical limit where the above commutators become
Poisson brackets and $H$, $S^{\pm}$ are functions, a similar representation of
the  ${\cal U}_{\hbar}(sl_{2})$ generators was obtained in \cite{BAB1} from a
completely different approach.}
\begin{eqnarray}
\left[ H, S^{\pm} \right] & = & \pm 2 S^{\pm} \nonumber\\
 \left[ S^{+}, S^{-} \right] & = & \frac{q^{H}-q^{-H}}{2 \hbar}
\end{eqnarray}
Moreover, the co-product action on the above ${\cal U}_{\hbar}(sl_{2})$
generators follows on from their geometrical definition. The co-product
$\Delta$ is obtained simply by doubling the interval $[0, 1]$ to $[0, 2]$ the
first space in the tensor product being associated to $[0, 1]$ while the second
space is associated to $[1, 2]$. We get,
\begin{eqnarray}
\Delta(H) & = &\Delta \left( \int_{0}^{1} du\ h(u)  \right)\nonumber\\
& = &  \int_{0}^{2} du\ h(u)  \nonumber\\
& = &  \int_{0}^{1} du\ h(u) +  \int_{1}^{2} du\ h(u)\nonumber\\
& = &  H \otimes {\bf 1} + {\bf 1} \otimes H \nonumber\\
\Delta(S^{+}) & = &\Delta \left( \int_{0}^{1} du\  e^{+}(u)\ q^{\int_{u}^{1}
dv\ h(v) } \right)\nonumber\\
& = &  \int_{0}^{2} du\  e^{+}(u)\ q^{\int_{u}^{2} dv\ h(v) } \nonumber\\
& = & \int_{0}^{1} du\  e^{+}(u)\ q^{\int_{u}^{1} dv\ h(v) }\ q^{\int_{1}^{2}
dv\ h(v) } +   \int_{1}^{2} du\ e_{+}(u)\ q^{\int_{u}^{2} dv\ h(v) }
\nonumber\\
& = &  S^{+}\otimes q^{H} + {\bf 1} \otimes S^{+}
\label{eq:d1}
\end{eqnarray}
and similarly, $\Delta(S^{-}) = S^{-} \otimes {\bf 1} + q^{-H}\otimes  S^{-}$.
\\
{}From the general results of the previous section, the above co-product
formulas can also be obtained from the adjoint action of $\cal F$ on the
trivial co-product  ${\Delta}_{0}$. Indeed with,
\begin{eqnarray*}
{\cal F}^{(+)}_{12} & = & \overrightarrow{\exp}\ ( \hbar \int_{0}^{1} d{\omega}
\ r^{+}_{12} \left[\omega\right] )\\
{\cal F}^{(-)}_{12} & = & \overrightarrow{\exp}\ ( \hbar \int_{0}^{1} d{\omega}
\ r^{-}_{12} \left[\omega\right] )\\
r^{+} & = &  \frac{1}{2} h \otimes h + 2 e^{+} \otimes e^{-} \\
r^{-} & = &  -\frac{1}{2} h \otimes h - 2  e^{-}\otimes e^{+}
\end{eqnarray*}
we get,
\[{\cal F}^{(+)} = q^{\frac{1}{2} \int_{0}^{1}\hspace{-4.5pt}{du} \left( h(u)
\otimes \int_{0}^{u}\hspace{-6pt}{dv}\ h(v)\right) }\; \overrightarrow{\exp}\ (
2 \hbar\int_{0}^{1} \hspace{-6pt}{du} \left( S^{+}(u) \otimes
\int_{0}^{u}\hspace{-6pt}{dv}\; S^{-}(v) \right) ) \]  \\
\[{\cal F}^{(-)} =  q^{-\frac{1}{2}\int_{0}^{1}\hspace{-4.5pt}{du} \left(h(u)
\otimes \int_{0}^{u}\hspace{-4.5pt}{dv}\ h(v)\right) }\ \overrightarrow{\exp}\
( -2 \hbar\int_{0}^{1}\hspace{-6pt}{du} \left(\tilde{S}^{-}(u) \otimes
\int_{0}^{u}\hspace{-6pt}{dv}\; \tilde{S}^{+}(v) \right) ) \]
\[
\begin{array}{ll}
\  \mbox{where}\;& \tilde{S}^{+}(u) = {S}^{+}(u)q^{-H} \\
& \tilde{S}^{-}(u) = q^{H}{S}^{-}(u)
\end{array} \]
and the above formulas for ${\cal F}^{(\pm)}$ imply that for any element $a \in
{{\cal U}_{\hbar}(sl_{2})}$
$$
{\Delta} (a)\ =\ {\Delta}^{(+)}(a)\ =\ {\Delta}^{(-)}(a)
$$
the co-product $\Delta$ being defined above by eq. (\ref{eq:d1}) and with,
$$
{\Delta}^{(\pm)}(a)\ =\ ({\cal F}^{(\pm)}_{12})^{-1}\ \Delta_{0}(a)\ {\cal
F}^{(\pm)}_{12}
$$
This fact reflects the corresponding property for the universal quantum
$R$-matrix $\cal R$. Hence $H, S^{(\pm)}$ define the set of generators of
${\cal U}_{\hbar}(sl_{2})$ realized as a Hopf sub-algebra of ${\cal
U}(\widehat{sl_{2}})$ which acquires the structure of a quasi-triangular Hopf
algebra with co-product ${\Delta} = {\Delta}^{(\pm)}$ and universal quantum
$R$-matrix ${\cal R} \in {{\cal U}_{\hbar}(sl_{2})} \otimes {{\cal
U}_{\hbar}(sl_{2})}$ that agrees with the one obtain in \cite{Drin1}.\\
\\

{\bf 4. Conclusion}\\
\\
In this letter, we have sketched out the general construction of  the quantum
universal $R$-matrix and of its associated quantum algebra as functionals of
the corresponding classical $r$-matrix and Lie algebra. We have shown that our
approach applies to the $sl_{2}$ case. There, ${\cal U}_{\hbar}(sl_{2})$ is
realized as a Hopf sub-algebra of ${\cal U}(\widehat{sl_{2}})$ with deformed
co-product $\Delta$ obtained by the adjoint action of the universal elements
${\cal F}^{(\pm)} \in {\cal U}(\widehat{sl_{2}}) \otimes {\cal
U}(\widehat{sl_{2}})$ on the co-commutative co-product ${\Delta}_{0}$ induced
from the $sl_{2}$ Lie algebra.
It gives a representation of quantum algebras and of universal $R$-matrices in
terms of current algebras.\\

Among previous attempts to perform a universal quantization of Lie bi-algebras
and of their associated classical $r$-matrices, we shall quote the result of
Drinfel'd \cite{Drin1} and \cite{Drin2} valid for the triangular case where the
classical $r$-matrix is skew-symmetric. Our results are obtained with the only
condition that the classical $r$-matrix satisfies the classical Yang-Baxter
equation. Note that in contrast to the  approach in \cite{Drin1} and
\cite{Drin2}, the algebra $\cal A$ such that ${\cal R} \in {\cal A} \otimes
{\cal A}$ is not directly related to ${\cal U}({\cal G})$ but rather to  ${\cal
U}(\hat {\cal G})$. The relation between these two construction deserves to be
investigated further.
Let us also remark that no obstruction to quantization shows up in our
construction of the universal quantum $R$-matrix. However, this point should be
studied in the framework of
the important question of representation theory. From our result, there should
be a direct link between the representation theory of current algebras and of
quantum algebras. In particular, when going to the case of centrally extended
current algebras, the results of \cite{FM1} should  be recovered. \\
\\
\\

\end{document}